\documentclass[a4paper]{jpconf}

\usepackage{amsmath,amssymb,epsfig,fleqn,url,psboxit,color,here,graphics,graphicx,rotating,multirow,wrapfig}
\usepackage{bigstrut}
\usepackage{hyperref}
\hypersetup{
    colorlinks,
    citecolor=blue,
    filecolor=blue,
    linkcolor=blue,
    urlcolor=blue
}

\newcommand{\slim}{\mskip 1.5mu}              

\newcommand{\phiH}{\phi _h}
\newcommand{\phiS}{\phi _S}
\newcommand{\phiCS}{\varphi _{CS}}
\newcommand{\phiSDY}{\varphi _{S}}
\newcommand{\red}{\textcolor[rgb]{1.00,0.00,0.00}}
\newcommand{\blue}{\textcolor[rgb]{0.00,0.00,1.00}}
\begin{document}
\title{Transverse spin azimuthal asymmetries at COMPASS: SIDIS Multi\--D analysis \& Drell-Yan}

\author{Bakur Parsamyan\\ (on behalf of the COMPASS collaboration)}

\address{University of Turin and Torino Section of INFN\\
 Via P. Giuria 1, 10125 Torino, Italy }

\ead{bakur.parsamyan@cern.ch}

\begin{abstract}
COMPASS is a high-energy physics experiment operating on the M2 beam line at the SPS at CERN.
Using high energy muon and hadron beams the experiment covers broad range of physics aspects in the field
of the hadron structure and spectroscopy. One of the important objectives of the COMPASS experiment is the exploration
of transverse spin structure of the nucleon via study of spin (in)dependent azimuthal asymmetries with semi-inclusive deep
inelastic scattering (SIDIS) processes and starting from 2014 also with Drell-Yan (DY) reactions.
Experimental results obtained by COMPASS for azimuthal effects in SIDIS play an important role in the general understanding
of the three-dimensional nature of the nucleon. Giving access to the entire "twist-2" set of transverse momentum dependent
(TMD) parton distribution functions (PDFs) and fragmentation functions (FFs) COMPASS data trigger constant theoretical
interest and are being widely used in phenomenological analyses and global data fits. In particular, unique x-$Q^{2}$-z-pT
multidimensional results for transverse spin asymmetries recently obtained by COMPASS will serve as a direct and unprecedented
input for TMD $Q^{2}$-evolution related studies, one of the hottest topics in the field of spin-physics.	
In addition, measurement of the Sivers and all other azimuthal effects in polarized Drell-Yan at COMPASS will reveal another
side of the spin-puzzle providing a link between SIDIS and Drell-Yan branches. This will be a unique possibility to test
universality and key-features of TMD PDFs using essentially the same experimental setup and exploring the same kinematical domain. 	
In this review main focus will be given to the very recent
results obtained by the collaboration for multi-dimensional transverse spin asymmetries and to the physics aspects of
COMPASS polarized Drell-Yan program.
\end{abstract}
%
%
%
%
%
%
%
%
\section{Introduction}	
%
Detailed examination of azimuthal asymmetries arising in the SIDIS and Drell-Yan cross-sections is a powerful method
used to access TMD distribution functions of the nucleon. Within the LO QCD parton model approach the polarized nucleon
is described by six time reversal even and two
time reversal odd twist-two TMD PDFs which within the scope of QCD-formalism are expected to be universal between different reactions
\footnote{QCD \textit{generalized universality}: time-reversal modified process-independence of TMD PDFs} \cite{Kotzinian:1994dv}--\cite{Mulders:1995dh}.
In past decades spin-phenomena turned to be one of the hottest topics of modern science and thus, measurements and following study
of the spin dependent and \emph{unpolarized} azimuthal effects in
SIDIS and Drell-Yan became a priority direction in experimental and theoretical high-energy physics.
The ultimate goal is to measure experimentally with high precision all possible spin-effects with both SIDIS and Drell-Yan
reactions at different energies and
perform global multi-differential analysis of obtained results to extract all spin-dependent distribution functions.

\begin{wrapfigure}{r}{0.45\textwidth}
  \begin{center}
    \includegraphics[width=0.45\textwidth]{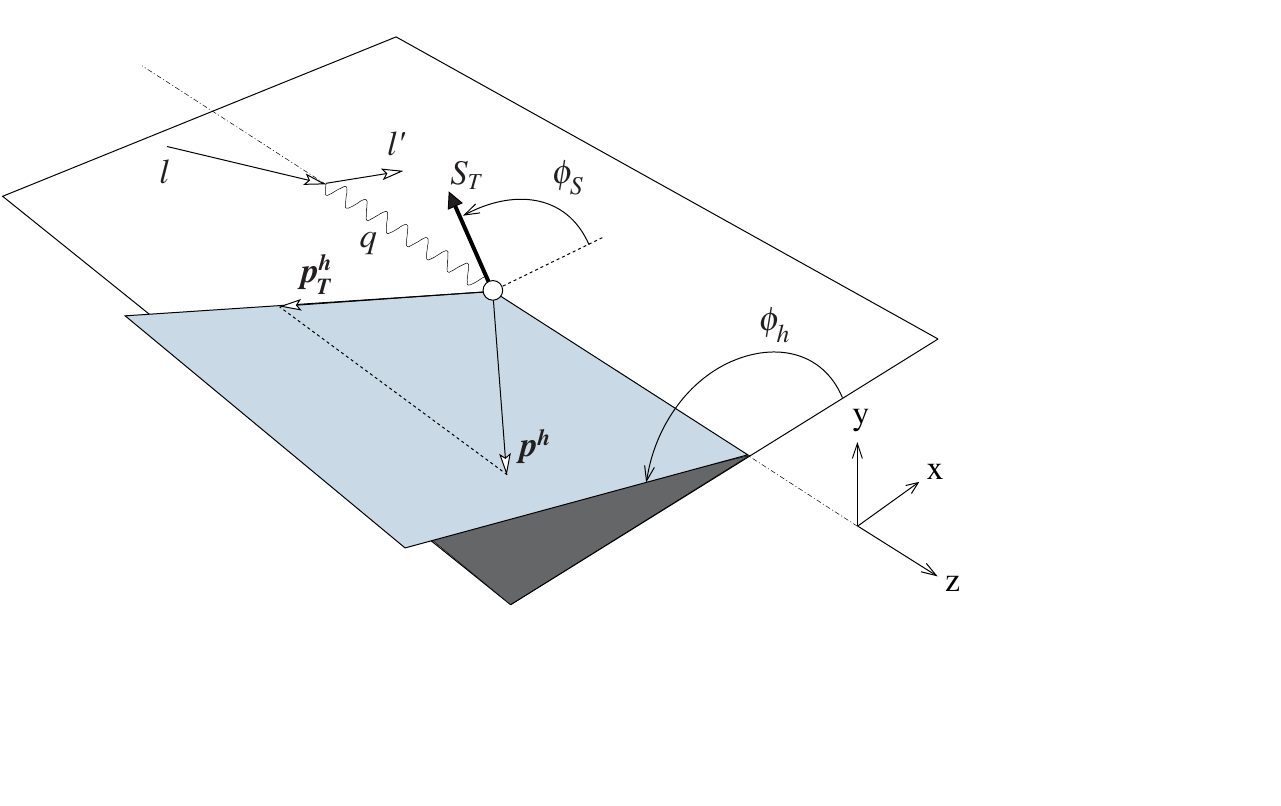}
  \end{center}
\caption{SIDIS process framework. Definition of azimuthal angles $\phiH$ and $\phiS$.}
  \label{fig:SIDISangles}
\end{wrapfigure}

Using standard notations the cross-section expression for the lepton off transversely polarized nucleon SIDIS processes
can be written in a following model-independent way \cite{Kotzinian:1994dv}--\cite{Diehl:2005pc}:
{\small
\begin{eqnarray}\nonumber
&& \hspace*{-1.8cm}\frac{{d\sigma }}{{dxdydzp_{T}^{h}dp_{T}^{h}d{\phiH}d\phiS }} = 2\left[ {\frac{\alpha }{{xy{Q^2}}}\frac{{{y^2}}}{{2\left( {1 - \varepsilon } \right)}}\left( {1 + \frac{{{\gamma ^2}}}{{2x}}} \right)} \right]\left( {{F_{UU,T}} + \varepsilon {F_{UU,L}}} \right) \\
%
%
&&\hspace*{-1.8cm} \times\Bigg\{ 1 + \sqrt {2\varepsilon \left( {1 + \varepsilon } \right)} \blue{A_{UU}^{\cos {\phi _h}}}\cos {\phiH} + \varepsilon \red{A_{UU}^{\cos 2{\phi _h}}}\cos \left( {2{\phiH}} \right) + \lambda \sqrt {2\varepsilon \left( {1 - \varepsilon } \right)} \blue{A_{LU}^{\sin {\phi _h}}}\sin {\phiH}\\ \nonumber
&&\hspace*{-1.5cm}+\,{{S}_{T}}\Big[\red{A_{UT}^{\sin \left( {{\phiH} - {\phiS}} \right)}}\sin \left( {{\phiH} - {\phiS}} \right) + \varepsilon \left(\red{A_{UT}^{\sin \left( {{\phiH} + {\phiS}} \right)}}\sin \left( {{\phiH} + {\phiS}} \right) + \red{A_{UT}^{\sin \left( {3{\phiH} - {\phiS}} \right)}}\sin \left( {3{\phiH} - {\phiS}} \right)\right)\\ \nonumber
&&\hspace*{-0.3cm}+\,\sqrt {2\varepsilon \left( {1 + \varepsilon } \right)} \left(\blue{A_{UT}^{\sin {\phiS}}}\sin {\phiS} + \blue{A_{UT}^{\sin \left( {2{\phiH} - {\phiS}} \right)}}\sin \left( {2{\phiH} - {\phiS}} \right)\right)\Big]\\ \nonumber
&&\hspace*{-1.5cm}+\,{{S}_{T}}\lambda \Big[\sqrt {\left( {1 - {\varepsilon ^2}} \right)} \red{A_{LT}^{\cos \left( {{\phiH} - {\phiS}} \right)}}\cos \left( {{\phiH} - {\phiS}} \right)
+\,\sqrt {2\varepsilon \left( {1 - \varepsilon } \right)} \left(\blue{A_{LT}^{\cos {\phiS}}}\cos {\phiS} + \blue{A_{LT}^{\cos \left( {2{\phiH} - {\phiS}} \right)}}\cos \left( {2{\phiH} - {\phiS}} \right)\right)\Big]\Bigg\},
\label{eq:SIDIS}
\end{eqnarray}
}
with ratio of longitudinal and transverse photon fluxes given as $\varepsilon = (1-y -\frac{1}{4}\slim \gamma^2 y^2)/(1-y +\frac{1}{2}\slim y^2 +\frac{1}{4}\slim \gamma^2
y^2)$ and $\gamma = 2 M x/Q$.
 Target transverse polarization (${S}_{T}$)
dependent part \footnote{in reality polarization vector has a small longitudinal component w.r.t. the $\gamma^*$ momenta which at COMPASS kinematics leads
to slight deviations of discussed TSAs (for details see \cite{Parsamyan:2013ug})}
of this general expression contains eight azimuthal modulations
in the $\phi_h$ and $\phi_S$ (azimuthal angles of the produced hadron and of the nucleon spin, correspondingly (see Figure~\ref{fig:SIDISangles})).
Each modulation leads to a $A_{BT}^{w_i(\phiH, \phiS)}$ Transverse-Spin-dependent Asymmetry (TSA) defined as a ratio of
the corresponding structure function $F_{BT}^{w_i(\phiH,\phiS)}$ to the \emph{unpolarized} one
${F_{UU}}={{F_{UU,T}} + \varepsilon {F_{UU,L}}}$. Here the superscript of the asymmetry
indicates respective modulation, while "U"-unpolarized, "L"-longitudinal and
"T"-transverse subscripts denote beam (B) and target (T) polarizations. Five amplitudes are called Single-Spin
Asymmetries (SSA) since they depend only on ${S}_{T}$. The other three depend both on ${S}_{T}$ and $\lambda$ beam longitudinal polarization and are known as
Double-Spin Asymmetries (DSA).
%
%
%

In the QCD parton model approach four out of eight transverse spin asymmetries have Leading Order (LO) interpretation
in terms of convolutions of twist-two transverse-momentum-dependent parton
distribution functions and fragmentation functions \cite{Kotzinian:1994dv}--\cite{Mulders:1995dh}.
The first two LO asymmetries: $A_{UT}^{sin(\phi_h-\phi_S)}$ "Sivers" and $A_{UT}^{sin(\phi_h+\phi_S)}$ "Collins" effects \cite{Adolph:2012sn,Adolph:2012sp}
are the most studied ones. Corresponding structure functions are given as convolutions of
 $f_{1T}^{\perp q}$ Sivers PDF with $D_{1q}^h$ ordinary FF and $h_{1}^{q}$ "transversity" PDF with the $H_{1q}^{\perp h}$ Collins FF, respectively.
 The other two LO terms give access to another pair of twist-two TMD PDFs: $A_{UT}^{\sin(3\phiH -\phiS )}$ SSA is related to $h_{1T}^{\perp\,q}$ ("pretzelosity") PDF
 \cite{Parsamyan:2014uda}--\cite{Parsamyan:2007ju}) and $A_{LT}^{\cos (\phiH -\phiS )}$ DSA to the $g_{1T}^q$ ("worm-gear") distribution function
 \cite{Parsamyan:2014uda}--\cite{Parsamyan:2007ju,Kotzinian:2006dw,Anselmino:2006yc}).
{\small
\begin{align}\label{eq:LO_as}
&\red{A_{UT}^{\sin (\phiH -\phiS )}} \propto f_{1T}^{\bot q} \otimes
D_{1q}^h,\ \
\red{A_{UT}^{\sin (\phiH +\phiS )}} \propto h_1^q \otimes H_{1q}^{\bot
h},  \\
&\red{A_{UT}^{\sin (3\phiH
-\phiS )}} \propto h_{1T}^{\bot q} \otimes H_{1q}^{\bot
h},\ \red{A_{LT}^{\cos (\phiH -\phiS )}} \propto g_{1T}^q \otimes
D_{1q}^h.\nonumber
\end{align}
}
Remaining four asymmetries 
are so-called "higher-twist" effects\footnote{in equations \ref{eq:SIDIS} and \ref{eq:DY} the twist-2 amplitudes are marked in red and
higher-twist ones in blue}. Linked sub-leading $Q^{-1}$-order structure functions contain
terms given as various mixtures of twist-two and quark-gluon correlation induced twist-three parton distribution
and fragmentation functions \cite{Bacchetta:2006tn,Mao:2014aoa,Mao:2014fma}. However, applying wildly adopted
"Wandzura-Wilczek approximation" this higher twist expressions can be simplified
to twist-two level (see \cite{Bacchetta:2006tn,Mulders:1995dh} for more details):
{\small
\begin{align}\label{eq:NLO_as}
&\blue{A_{UT}^{\sin (\phiS )}} \propto {Q}^{-1}({h_1^q \otimes
H_{1q}^{\bot h} +f_{1T}^{\bot q} \otimes D_{1q}^h }),\
\blue{A_{UT}^{\sin (2\phiH -\phiS )}} \propto
{Q}^{-1}({h_{1T}^{\bot q} \otimes H_{1q}^{\bot h}
+f_{1T}^{\bot q} \otimes D_{1q}^h }),\ \\
&\blue{A_{LT}^{\cos (\phiS )}} \propto {Q}^{-1}(g_{1T}^q \otimes
D_{1q}^h),\ \
\blue{A_{LT}^{\cos (2\phiH -\phiS )}} \propto {Q}^{-1}
(g_{1T}^q \otimes D_{1q}^h).\nonumber
\end{align}
}
The whole set of eight SIDIS TSAs has been extracted from COMPASS transversely polarized deuteron and proton data
(See \cite{Adolph:2012sn}--\cite{Parsamyan:2007ju} and references therein).
%

\begin{wrapfigure}{r}{0.45\textwidth}
  \begin{center}
    \includegraphics[width=0.45\textwidth]{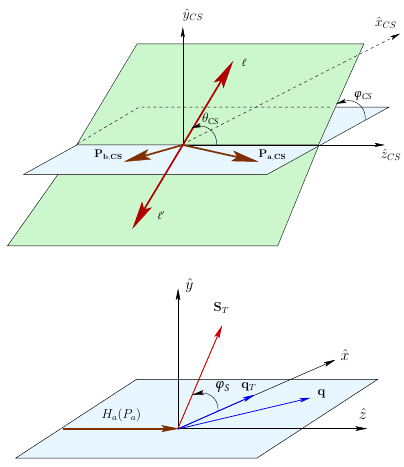}
  \end{center}
\caption{Drell-Yan process framework. Definition of azimuthal angles $\phiCS$ and $\phiSDY$.}
  \label{fig:DYangles}
\end{wrapfigure}

Applying similar notations, model-independent single-polarized ($\pi N^\uparrow$) Drell-Yan cross-section at leading order can be written
in the following way \cite{Gautheron:2010wva}:
{\small
\begin{eqnarray}
  &&\hspace*{-1.0cm}\frac{{d{\sigma ^{LO}}}}{{d\Omega }} = \frac{{\alpha _{em}^2}}{{F{q^2}}}F_U^1 \left\{ {1 + {{\cos }^2}\theta  + {{\sin }^2}\theta \red{A_U^{\cos 2{\varphi _{CS}}}}\cos 2{\varphi _{CS}}} \right. +
  {S_T}\left[ {\left( {1 + {{\cos }^2}\theta } \right)\red{A_T^{\sin {\varphi _S}}}\sin {\varphi _S}} \right.\hfill \\ \nonumber
  &&\hspace*{+1.9cm}{\text{   }}  + {\sin ^2}\theta \left(\red{A_T^{\sin \left( {2{\varphi _{CS}} + {\varphi _S}} \right)}}\sin \left( {2{\varphi _{CS}} + {\varphi _S}} \right)
   + \left. {\left. { \red{A_T^{\sin \left( {2{\varphi _{CS}} - {\varphi _S}} \right)}}\sin \left( {2{\varphi _{CS}} - {\varphi _S}} \right)} \right]}\right) \right\}.
\label{eq:DY}
\end{eqnarray}
}
where angular variables are defined in Collins-Soper and target rest frames (see Figure~\ref{fig:DYangles}).
Similarly to the SIDIS case, the superscript of the asymmetry indicates the corresponding modulation, while "U", "L" and "T" subscripts mark the target polarization.
As one can see, in the Drell-Yan cross-section only one unpolarized and three target transverse spin dependent azimuthal modulations arise at leading order.
Within the same QCD parton model approach, Drell-Yan TSAs are also interpreted in terms of TMD PDFs.
In this case the asymmetries are related to the convolution of two TMD PDFs: one of the beam and one of the target hadron.
Quoting only the target nucleon PDFs: the $A_{T}^{\sin \varphi _s }$, $A_{T}^{\sin
(2\varphi _{CS} -\varphi _s )}$ and $A_{T}^{\sin(2\varphi _{CS} +\varphi _s )}$ give access to the
"Sivers" $f_{1T}^{\perp\,q}$, "transversity" $h_1^q$ and "pretzelosity" $h_{1T}^{\perp\,q}$, distribution functions, respectively.
In accordance with QCD \textit{generalized universality} principle, TMD PDFs accessed
via azimuthal asymmetries both in SIDIS and Drell-Yan reactions are expected to be the same.
Thus, first ever polarized Drell-Yan data collected in 2015 at COMPASS contains information which is
intriguingly complementary to the previously accumulated SIDIS results.
Using essentially same experimental setup COMPASS collaboration took an unprecedented opportunity
to access TMD PDFs via two mechanisms and test their universality and key features
as, for instance, the predicted sign-change of Sivers and Boer-Mulders PDFs.
In Table.~\ref{tab:PDFs} nucleon TMD PDFs and relative SIDIS and DY asymmetries are listed.

In general, TSAs being convolutions of different TMD functions are known to be complex objects which \textit{a priori}
depend on the choice of multidimensional kinematical ranges. Thus, ideally, asymmetries
have to be extracted as multi-differential functions of kinematical variables in order to reveal the most complete
multivariate dependence. In practice, available experimental data often is too limited for such an ambitious approach.
In order to investigate dependence of the asymmetries on some specific kinematic variable one is forced to integrate over all the others,
simplifying the task to one-dimensional case.

Presently, one of the hottest topics in the field of spin-physics, which requires at least two-dimensional analysis,
is the study of TMD evolution of various PDFs and FFs and related asymmetries.
Attempting to describe available experimental observations and make predictions for the future
ones, different models predict from small up to quite large $\sim1/Q^2$ suppression of the QCD-evolution
effects \cite{Aybat:2011ta,Echevarria:2014xaa,Sun:2013hua}. Additional experimental measurements exploring different
$Q^2$ domains for fixed $x$-range are necessary to further constrain the theoretical models.
The work described in this review is a unique and first ever attempt to explore the multivariate
kinematical behaviour of TSAs.
For this purpose COMPASS experimental data was split into five different $Q^2$ ranges giving an opportunity to study
asymmetries as a function of $Q^2$ at fixed bins of $x$. Additional variation of $z$ and $p_T$ cuts
deeper explores multi-dimensional dependences of the TSAs and their TMD constituents.
{\small
\begin{table}[h]
\centering
\caption{\label{tab:PDFs}Nucleon TMD PDFs accessed via SIDIS and Drell-Yan TSAs.}
\vspace*{0.2cm}
\begin{tabular}{@{}ccc@{}} 
  \br
  SIDIS $\ell^\rightarrow N^\uparrow$ & TMD PDF & DY $\pi N^\uparrow$ (LO)\bigstrut\\
  \mr
  \red{$A_{UU}^{\cos 2\phi _h}$}, \blue{$A_{UU}^{\cos \phi _h}$} & $h_{1}^{\bot q}$& \red{$A_{U}^{\cos 2\varphi _{CS}}$} \bigstrut\\
  \mr
  \red{$A_{UT}^{\sin (\phi _h -\phi _s )}$}, \blue{$A_{UT}^{\sin\phi _s}$}, \blue{$A_{UT}^{\sin (2\phi _h -\phi _s )}$} & $f_{1T}^{\bot q}$& \red{$A_{T}^{\sin \varphi _{S}}$} \bigstrut\\
  \mr
  \red{$A_{UT}^{\sin (\phi _h +\phi _s -\pi)}$}, \blue{$A_{UT}^{\sin\phi _s}$} & $h_{1}^{q}$& \red{$A_{T}^{\sin (2\varphi _{CS} -\varphi _{S} )}$} \bigstrut\\
  \mr
  \red{$A_{UT}^{\sin (3\phi _h -\phi _s )}$}, \blue{$A_{UT}^{\sin (2\phi _h -\phi _s )}$} & $h_{1T}^{\bot q}$& \red{$A_{T}^{\sin (2\varphi _{CS} +\varphi _{S} )}$} \bigstrut\\
  \mr
  \red{$A_{LT}^{\cos (\phi _h -\phi _s )}$}, \blue{$A_{LT}^{\cos\phi _s}$}, \blue{$A_{LT}^{\cos (2\phi _h -\phi _s )}$} & $g_{1T}^{ q}$& \red{double-polarized DY} \bigstrut\\
 \br
\end{tabular}
\end {table}
}
\section{Multidimensional analysis of TSAs}	
During it's "phase-I" in 2002-2010 COMPASS has made series of SIDIS TSA measurements using 160 GeV/c
longitudinally polarized muon beam and transversely polarized $^6LiD$ and $NH_3$ targets (See \cite{Adolph:2012sn}--\cite{Parsamyan:2007ju} and references therein).
Within "phase-II" new measurements for TSAs, but this time with Drell-Yan reaction took place
in 2015 with 190 GeV/c $\pi^-$ beam and transversely polarized polarized $NH_3$-target \cite{Parsamyan:2015cfa,Gautheron:2010wva}.

Very soon, both sets of COMPASS results from SIDIS and Drell-Yan will become a subject of global phenomenological analyses.
For this purpose the best option is to explore SIDIS data in a more differential way extracting the asymmetries in the same four $Q^2$
regions which were selected for the COMPASS Drell-Yan measurement program \cite{Parsamyan:2015cfa,Gautheron:2010wva}:
$Q^{2}/(GeV/c)^2$ $\in$ $[1;4],[4;6.25],[6.25;16],[16;81]$.
COMPASS preliminary results with this selection have been already presented in \cite{Parsamyan:2014uda,Parsamyan:2015cfa}
while current review is dedicated to more recent $x$-$z$-$p_T$-$Q^2$ multi-dimensional extractions of TSAs \cite{Parsamyan:2015dfa}.

The analysis was carried out on COMPASS data collected in 2010 with transversely polarized proton data. General event
selection procedure and asymmetry extraction as well as systematic uncertainty evaluation techniques applied for this
analysis were identical to those used for recent COMPASS results on Collins, Sivers and other TSAs \cite{Adolph:2012sn}--\cite{Parsamyan:2007ju}.

The whole set of target transverse spin dependent asymmetries were extracted simultaneously from the fit using extended
unbinned maximum likelihood method. Obtained "raw" amplitudes have been then corrected for average depolarization factors from equation \ref{eq:depol} ($\varepsilon$-depending
factors in equation \ref{eq:SIDIS} standing in front of the amplitudes), dilution factor and target and beam (only
DSAs) polarizations evaluated in the given kinematical bin \cite{Adolph:2012sn}--\cite{Parsamyan:2007ju}.
{\small
\begin{eqnarray}\label{eq:depol}
    && D^{\sin(\phiH -\phiS )}(y) \cong 1,\;\;
    D^{\cos(\phiH -\phiS )}(y) = \sqrt {\left( {1 - \varepsilon^{2} } \right)} \approx \frac {y(2-y)} {1+(1-y)^2},\nonumber\\
    && D^{\sin(\phiH +\phiS )}(y) = D^{\sin(3\phiH -\phiS )}(y)
 = \varepsilon \approx \frac {2(1-y)} {1+(1-y)^2}, \\
    && D^{\sin(2\phiH -\phiS )}(y) = D^{\sin(\phiS )}(y) = \sqrt {2\varepsilon \left( {1 + \varepsilon } \right)} \approx \frac
{2(2-y)\sqrt{1-y}} {1+(1-y)^2}, \nonumber \\
    &&D^{\cos(2\phiH -\phiS )}(y) = D^{\cos(\phiS )}(y) = \sqrt {2\varepsilon \left( {1 - \varepsilon } \right)} \approx \frac
{2y\sqrt{1-y}}
    {1+(1-y)^2}.\nonumber
\end{eqnarray}
}

Primary sample is defined by the following standard DIS cuts: $Q^2>1$ $(GeV/c)^2$, $0.003<x<0.7$ and $0.1 <y < 0.9$ and
two more \textit{hadronic} selections: $p_T>0.1$ GeV/c and $z>0.1$.

\begin{figure}[h]
\includegraphics[width=0.5\textwidth]{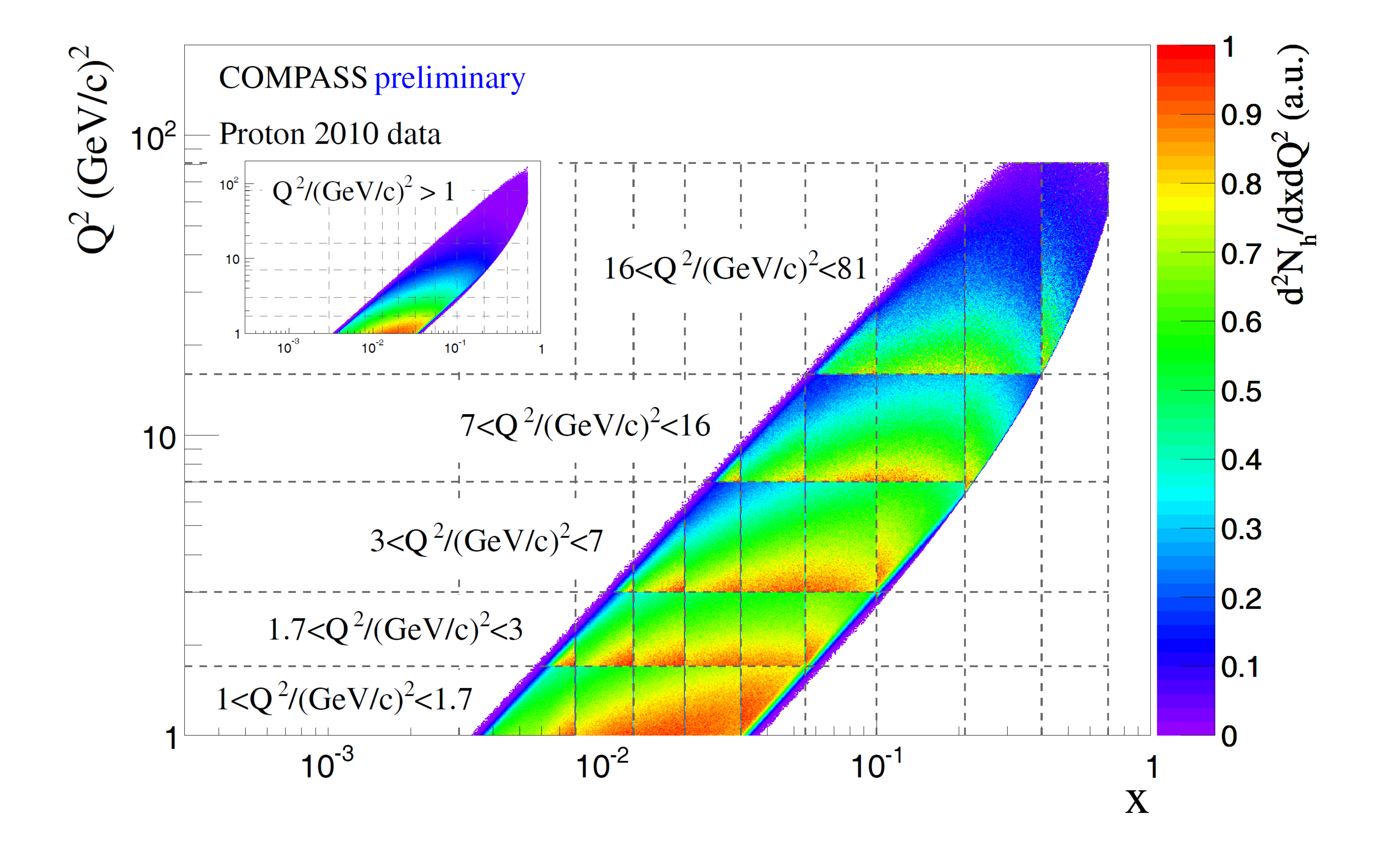}
\includegraphics[width=0.5\textwidth]{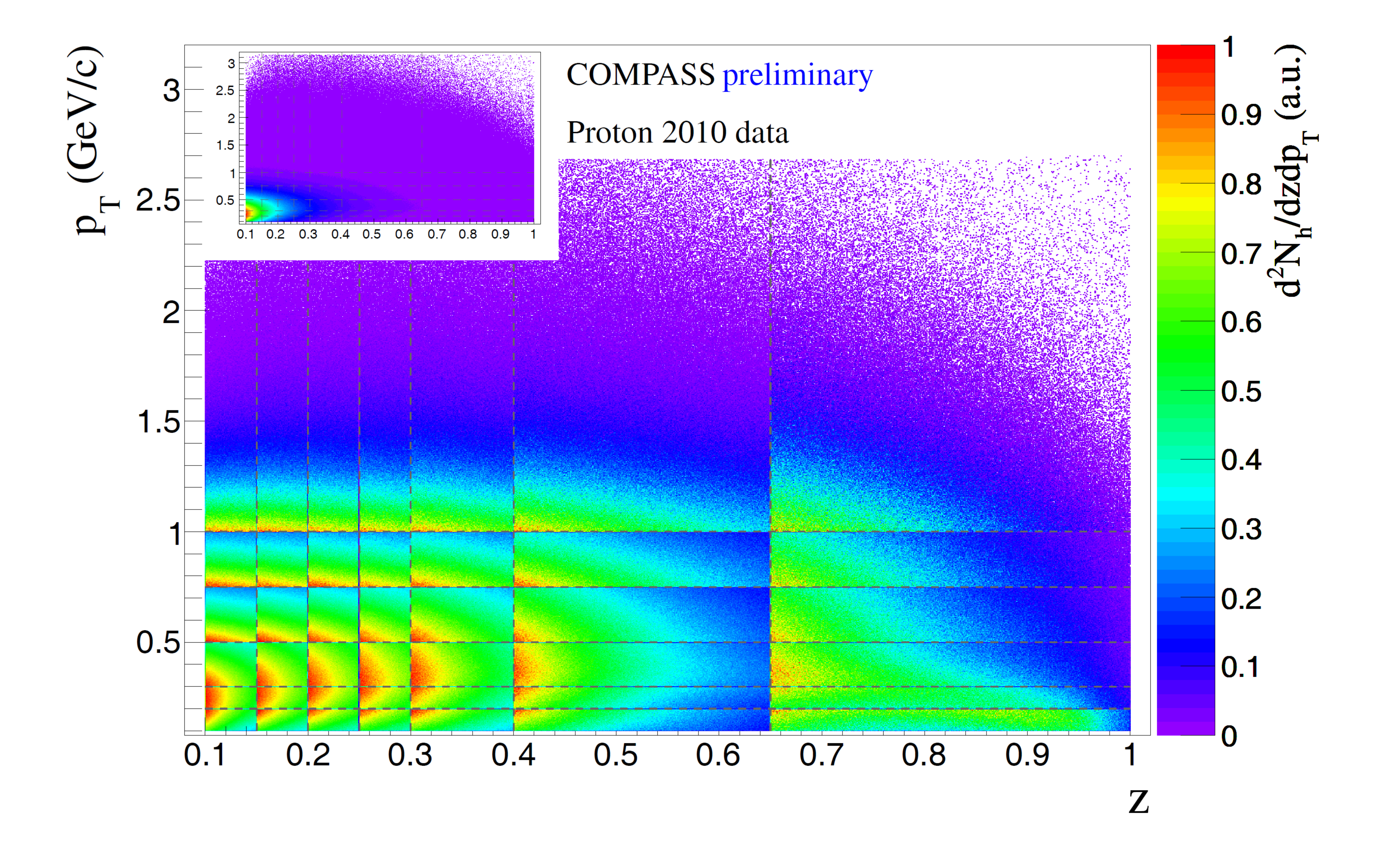}
\caption{COMPASS $x:Q^2$ (left) and $z$:$p_T$ (right) phase space coverage. \label{fig:f1}}
\end{figure}

In order to study possible $Q^2$-dependences of TSAs, the $x$:$Q^2$ phase-space covered by COMPASS experimental
data has been divided into $5\times9$ two-dimensional grid (see left plot in Figure~\ref{fig:f1}).
Selected five $Q^2$-ranges are the following ones: $Q^{2}/(GeV/c)^2$ $\in$ $[1;1.7],[1.7;3],[3;7],[7;16],[16;81]$.
In addition, each of this samples has been divided into five $z$ and five $p_T$ (GeV/c) sub-ranges defined as follows:\\
$z>0.1$, $z>0.2$, $0.1<z<0.2$, $0.2<z<0.4$ and $0.4<z<1.0$\\
$p_T>0.1$, $0.1<p_T<0.75$, $0.1<p_T<0.3$, $0.3<p_T<0.75$ and $p_T>0.75$.
Using various combinations of aforementioned cuts and sub-ranges, asymmetries have been
extracted for following "3D" and "4D" configurations: 1) $x$-dependence in $Q^2$-$z$ and $Q^2$-$p_T$ grids.
2) $Q^2$-dependence in $x$-$z$ and $x$-$p_T$ grids. 3) $Q^2$- (or $x$-) dependence in $x$-$p_T$ (or $Q^2$-$p_T$)
grids for different choices of $z$-cuts.
Another approach was used to examine $z$- and $p_T$-dependences in different $x$-ranges.
 In this study the two-dimensional $z$:$p_T$ phase-space has been divided into $7\times6$ grid
as it is demonstrated in right plot in Figure~\ref{fig:f1}.
Selecting three $x$-ranges: $0.003<x<0.7$, $0.003<x<0.032$, $0.032<x<0.7$ asymmetries
have been extracted in "3D: $x$-$z$-$p_T$" grid.
In the next section COMPASS preliminary results obtained for multi-dimensional
target transverse spin dependent azimuthal asymmetries are discussed.
\section{Results}
\begin{figure}[H]
\center
\includegraphics[width=1\textwidth]{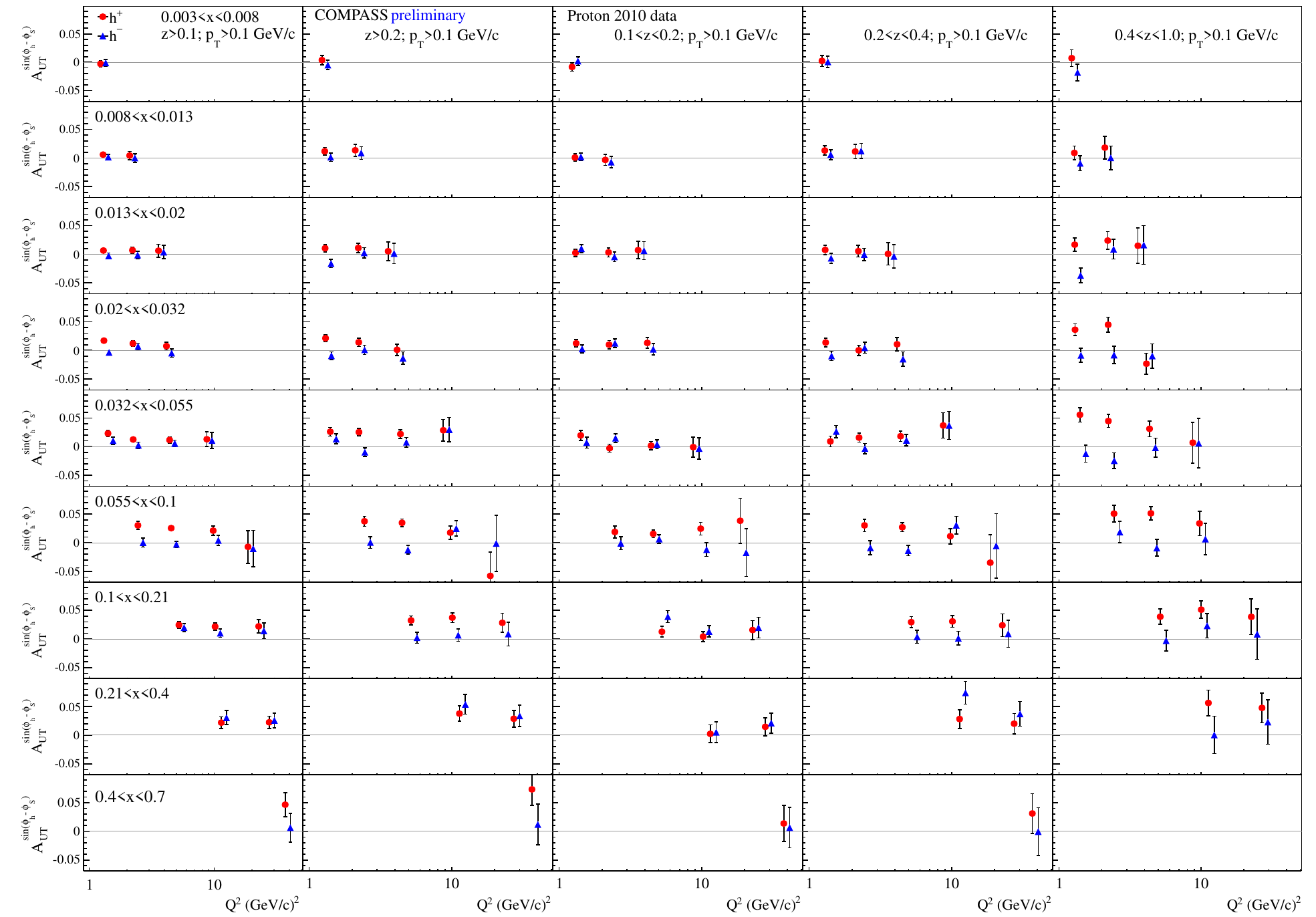}
\caption{Sivers asymmetry in "3D": $Q^2$-$p_T$-$x$ (top) and $x$-$z$-$Q^2$ (bottom). \label{fig:f2}}
\end{figure}

\footnote{Results discussed in this section have been first presented at the SPIN-2014 conference \cite{Parsamyan:2015dfa},
see also \cite{Parsamyan:2015myq},\cite{Parsamyan:QCDEV15}.}
Since it is difficult to make a detailed summary of the whole multidimensional variety of TSAs obtained by COMPASS in a short review format,
only selected results will be quoted in the following.
Extracted "3D: $x$-$z$-$Q^2$" Sivers effect is presented in the
Figure~\ref{fig:f2}. The results shown at the plot illustrate the $Q^2$-dependence of the asymmetry and
thus serve as a direct input for TMD-evolution related studies. In fact, in
several x-bins there are some hints for possible decreasing $Q^2$-dependence for positive hadrons which becomes more
evident at large $z$. In the meantime, Sivers asymmetry on positive hadrons tends to increase with both $z$ and $p_T$.
For negative hadrons, effect is compatible with zero except some indications for a positive signal at relatively large $x$ and $Q^2$
and negative effect at low $x$.

\begin{figure}[H]
\center
\includegraphics[width=1\textwidth]{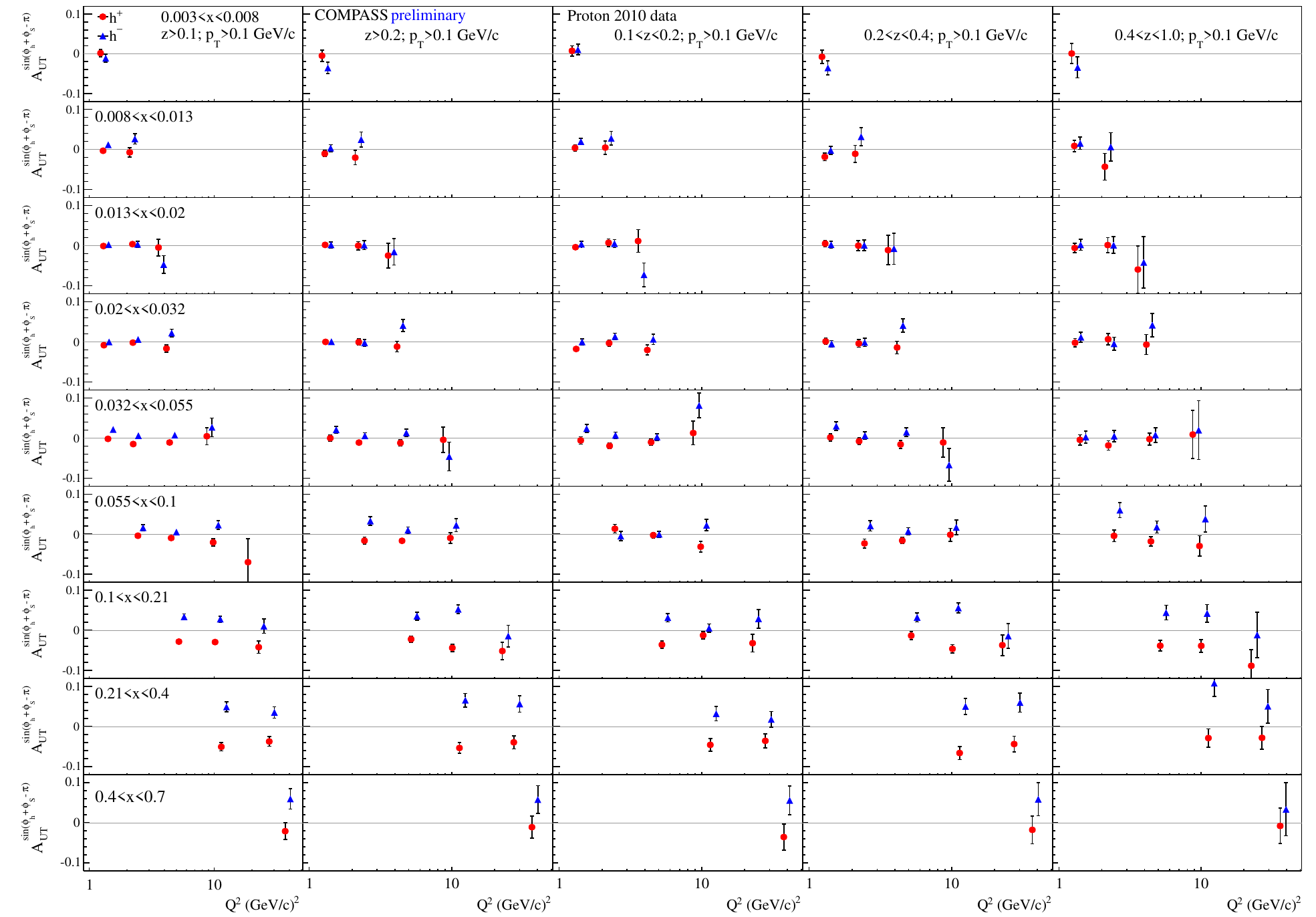}
\includegraphics[width=1\textwidth]{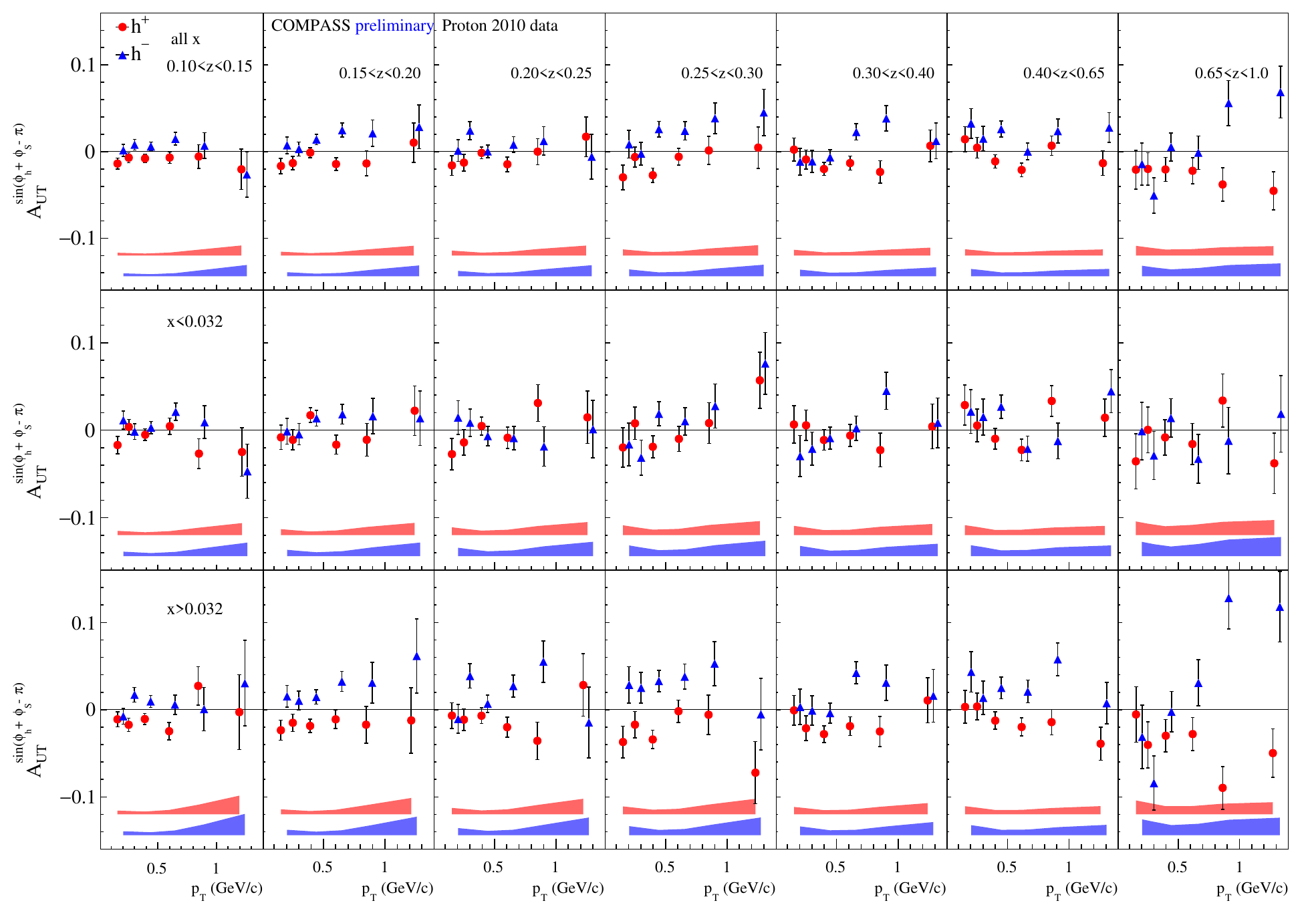}
\caption{Collins asymmetry in "3D": $x$-$z$-$Q^2$" (top) and "$x$-$z$-$p_T$" (bottom). \label{fig:f3}}
\end{figure}

In Figure~\ref{fig:f3} Collins asymmetry for the same "3D: $x$-$z$-$Q^2$"-configuration and for "3D: $x$-$z$-$p_T$" is shown.
Clear "mirrored" behaviour for positive and
negative hadron amplitudes is being observed in most of the bins. Amplitudes tend to increase in absolute value with
both $z$ and $p_T$. There are no clear indications for $Q^2$-dependence of Collins effect.

\begin{figure}[H]
\center
\includegraphics[width=1\textwidth]{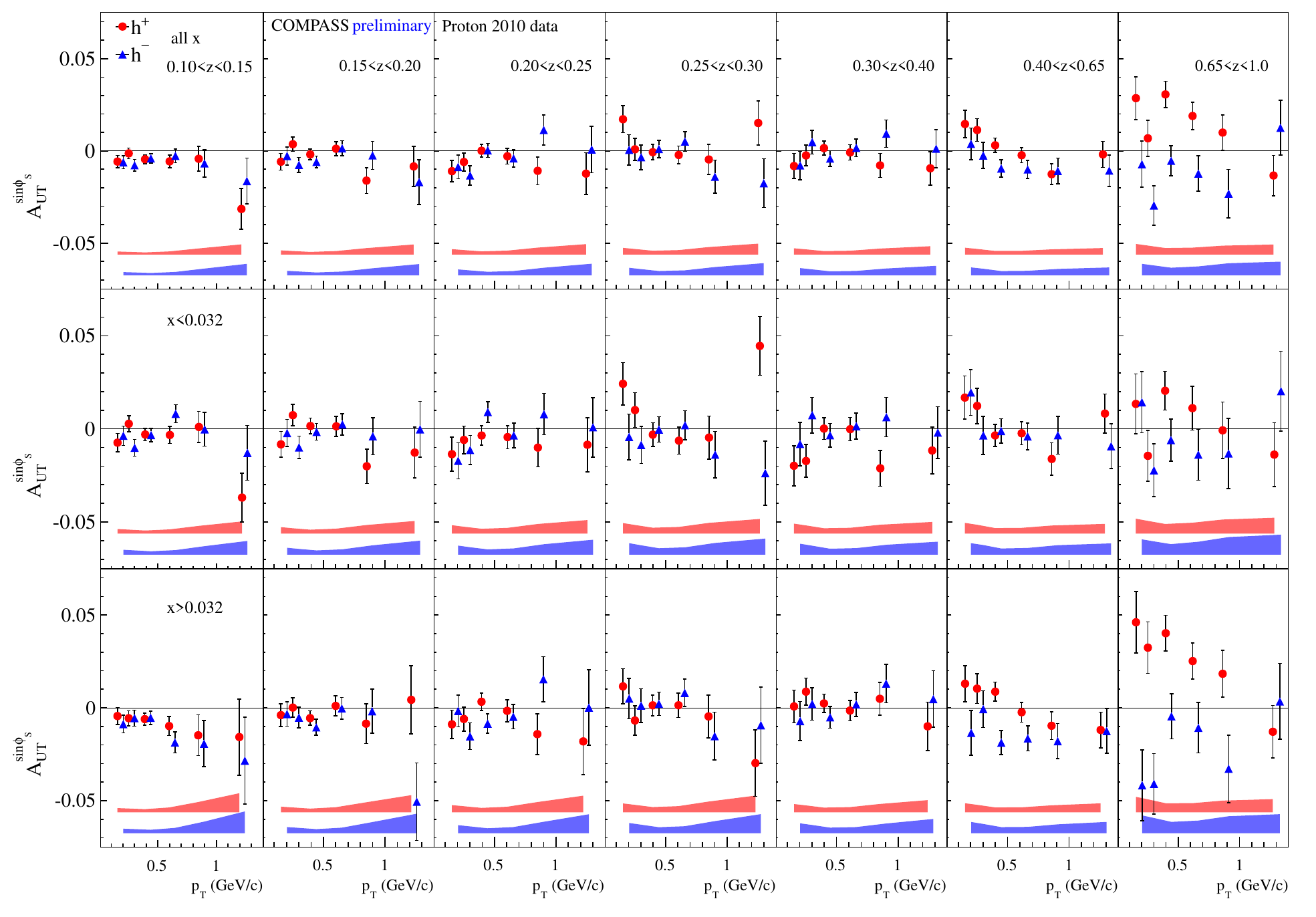}
\includegraphics[width=1\textwidth]{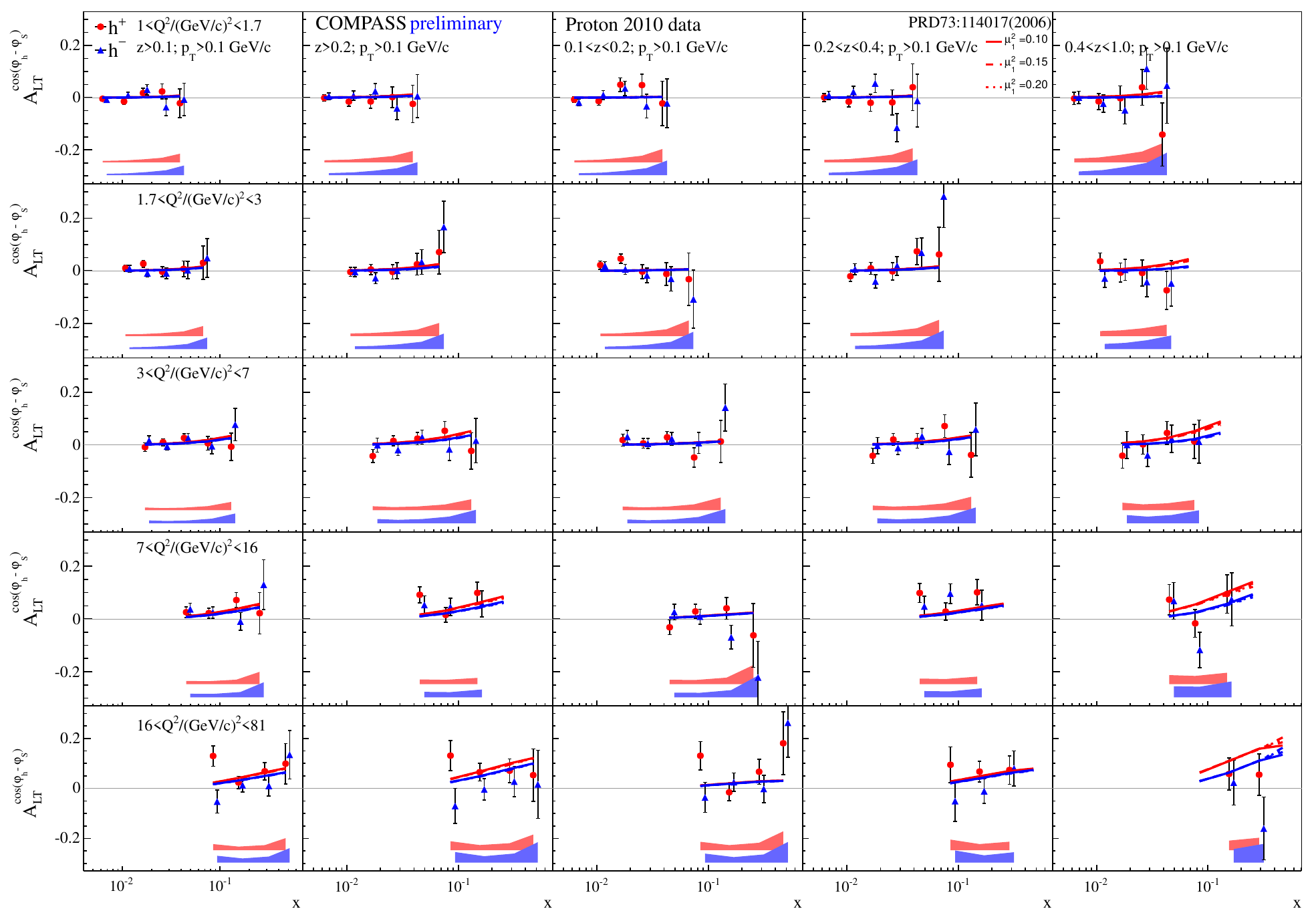}
\caption{Top: $A_{UT}^{\sin (\phi _s )}$ asymmetry in "3D" ($x$-$z$-$p_T$). Bottom: $A_{LT}^{\cos
(\phiH -\phiS )}$ in "3D" ($Q^2$-$z$-$x$) superimposed with theoretical predictions from [13].
\label{fig:f4}}
\end{figure}

Last SSA which is found to be non-zero at COMPASS is the higher-twist $A_{UT}^{\sin (\phi _s )}$ term which is presented in
Figure~\ref{fig:f4} (top) in "3D: $x$-$z$-$p_T$" configuration. Here the most interesting is the large $z$-range were
amplitude is measured to be sizable and non zero both for positive and negative hadrons.

The bottom plot in the Figure~\ref{fig:f4} is dedicated to the $A_{LT}^{\cos (\phiH -\phiS )}$ LO DSA explored in "3D:
$Q^2$-$z$-$x$" grid and superimposed with the theoretical curves from \cite{Kotzinian:2006dw}. This is the only DSA
which appears to be non-zero at COMPASS and the last TSA for which a statistically significant signal has been
detected. Remaining four asymmetries are found to be small or compatible with zero within available statistical
accuracy which is in agreement with available predictions \cite{Mao:2014aoa,Mao:2014fma,Lefky:2014eia}.
\vspace*{-1.0cm}
\vspace*{-1.0cm}
\section{Conclusions}
COMPASS experiment has performed a first ever multidimensional extraction of the whole set of target transverse spin dependent azimuthal asymmetries
from polarized proton data.
Various multi-differential configurations has been tested exploring $x$:$Q^2$:$z$:$p_T$ kinematical phase-space.
Particular attention was given to the revelation of possible $Q^2$-dependence of
TSAs, serving a direct input to TMD-evolution related studies. Several interesting observations have been made studying
the results obtained for Sivers,
 Collins, $A_{LT}^{cos(\phi_h-\phi_S)}$ and $A_{UT}^{sin(\phi_S)}$ asymmetries. Other four asymmetries were found to be
compatible with zero within given statistical accuracy. Highly differential data set obtained for the eight TSAs,
combined with past and future relevant data obtained by other collaborations will give a unique opportunity to access
the whole set of TMD PDFs and test their multi-differential nature.
Also particularly interesting will be the future comparison with first ever polarized Drell-Yan data collected by COMPASS in 2015.
This unique opportunity to access nucleon spin-structure via two processes will be the first direct chance to test the universality
and key features of TMD PDFs sticking to the same $x$:$Q^2$ kinematical range.
\section*{References}\label{refs}
%

\begin{thebibliography}{99}
\providecommand{\eprint}[2][]{\href{http://www.arxiv.org/abs/#2}{{\tt
  arXiv:#2}}}
%
%



%
\bibitem{Kotzinian:1994dv}
  A.~Kotzinian,
  \emph{Nucl.\ Phys.\ B} {\bf 441}, 234 (1995)
  (\emph{Preprint} \eprint{hep-ph/9412283}).

%
\bibitem{Bacchetta:2006tn}
  A.~Bacchetta {\it et al.}
  \emph{JHEP} {\bf 0702}, 093 (2007)
  (\emph{Preprint} \eprint{hep-ph/0611265}).
%
\bibitem{Diehl:2005pc}
  M.~Diehl and S.~Sapeta,
  \emph{Eur.\ Phys.\ J.\ C} {\bf 41}, 515 (2005)
  (\emph{Preprint} \eprint{hep-ph/0503023}).
%
\bibitem{Mulders:1995dh}
  P.~J.~Mulders and R.~D.~Tangerman,
  \emph{Nucl.\ Phys.\ B} {\bf 484}, 538 (1997) 
  (\emph{Preprint} \eprint{hep-ph/9510301}).
%
%
\bibitem{Adolph:2012sn}
C.~Adolph {\it et al.}
  \emph{Phys.\ Lett.\ B} {\bf 717}, 376 (2012)
  (\emph{Preprint} \eprint{1205.5121}).
%
\bibitem{Adolph:2012sp}
C.~Adolph {\it et al.}
  \emph{Phys.\ Lett.\ B} {\bf 717}, 383 (2012)
  (\emph{Preprint} \eprint{1205.5122}).
%
%
\bibitem{Parsamyan:2014uda}
  B.~Parsamyan,
  \emph{EPJ Web Conf.}  {\bf 85} (2015) 02019
  (\emph{Preprint} \eprint{1411.1568}).
%
\bibitem{Parsamyan:2015dfa} B.~Parsamyan, TBP in \emph{Int. J. Mod. Phys. Conf. Ser.}
  (\emph{Preprint} \eprint{1504.01599}).
%
%
\bibitem{Parsamyan:2015cfa}
  B.~Parsamyan,
  TBP in \emph{Int. J. Mod. Phys. Conf. Ser.}
  (\emph{Preprint} \eprint{1504.01598}).
%
%
\bibitem{Parsamyan:2013ug}
  B.~Parsamyan,
  \emph{Phys.\ Part.\ Nucl.}  {\bf 45}, 158 (2014)
  (\emph{Preprint} \eprint{1301.6615}).
%
%
\bibitem{Parsamyan:2013fia}
  B.~Parsamyan,
  \emph{PoS DIS} {\bf 2013}, 231 (2013)
  (\emph{Preprint} \eprint{1307.0183}).
%
%
\bibitem{Parsamyan:2010se}
  B.~Parsamyan,
  \emph{J.\ Phys.\ Conf.\ Ser.}  {\bf 295}, 012046 (2011)
  (\emph{Preprint} \eprint{1012.0155}).
%
\bibitem{Parsamyan:2007ju}
  B.~Parsamyan,
 \emph{ Eur.\ Phys.\ J.\ ST} {\bf 162}, 89 (2008)
  (\emph{Preprint} \eprint{0709.3440}).
%
\bibitem{Kotzinian:2006dw}
  A.~Kotzinian {\it et al.}
  \emph{Phys.\ Rev.\ D} {\bf 73}, 114017 (2006)
  (\emph{Preprint} \eprint{hep-ph/0603194}).
%
\bibitem{Anselmino:2006yc}
  M.~Anselmino {\it et al.}
  \emph{Phys.\ Rev.\ D} {\bf 74}, 074015 (2006)
  (\emph{Preprint} \eprint{hep-ph/0608048}).
%
%
\bibitem{Gautheron:2010wva}
   F.~Gautheron {\it et al.}
  ``COMPASS-II Proposal''
  CERN-SPSC-2010-014.
%
%
\bibitem{Mao:2014aoa}
  W.~Mao, Z.~Lu and B.~Q.~Ma,
  \emph{Phys.\ Rev.\ D} {\bf 90} (2014) 1,  014048
  (\emph{Preprint} \eprint{1405.3876}).
%
%
\bibitem{Mao:2014fma}
  W.~Mao {\it et al.}
  \emph{Phys.\ Rev.\ D} {\bf 91} (2015) 3,  034029
  (\emph{Preprint} \eprint{1412.7390}).
%
\bibitem{Aybat:2011ta}
  S.~M.~Aybat, A.~Prokudin and T.~C.~Rogers,
  \emph{Phys.\ Rev.\ Lett.}  {\bf 108} (2012) 242003
  (\emph{Preprint} \eprint{1112.4423}).
%
\bibitem{Echevarria:2014xaa}
  M.~G.~Echevarria {\it et al.}
  \emph{Phys.\ Rev.\ D} {\bf 89}, 074013 (2014)
  (\emph{Preprint} \eprint{1401.5078}).
%
\bibitem{Sun:2013hua}
  P.~Sun and F.~Yuan,
  \emph{Phys.\ Rev.\ D} {\bf 88}, no. 11, 114012 (2013)
  (\emph{Preprint} \eprint{1308.5003}).
%
%
\bibitem{Parsamyan:2015myq}
  B.~Parsamyan,
  \emph{CIPANP2015 Conference Proceedings}
  (\emph{Preprint} \eprint{1511.09093}).
%
%
\bibitem{Parsamyan:QCDEV15}
  B.~Parsamyan,
  \emph{QCDEV2015 Conference Proceedings}
  %
  (\emph{Preprint} \eprint{1512.06590}).
%
%
\bibitem{Lefky:2014eia}
  C.~Lefky and A.~Prokudin,
  \emph{Phys.\ Rev.\ D} {\bf 91} (2015) 3,  034010
  (\emph{Preprint} \eprint{1411.0580}).
%


\end{thebibliography}
%

\end{document}